\documentclass[10pt,final,journal,twocolumn]{IEEEtran}

\ifCLASSINFOpdf

\else

\fi

\usepackage{cite}
\usepackage[cmex10]{amsmath}
\usepackage{amsmath}
\usepackage{algorithmic}
\usepackage{stfloats}
\usepackage{multicol,multienum}
\usepackage{multirow}
\usepackage[dvips]{graphicx}

\usepackage[tight,footnotesize]{subfigure}
\usepackage{wrapfig}
\usepackage{color}

 \usepackage{booktabs}
 \usepackage{multirow}

\usepackage{mathtools}
\usepackage{color}
\usepackage{float}
\usepackage{bm}

\usepackage{pifont}

\usepackage[ruled,linesnumbered]{algorithm2e}

\usepackage{bigfoot}

\begin{document}

\title{Enhancing Wireless Networks with Attention Mechanisms: Insights from Mobile Crowdsensing}

\author{Yaoqi~Yang,
Hongyang~Du,
Zehui~Xiong,~\IEEEmembership{Senior Member,~IEEE},
Dusit~Niyato,~\IEEEmembership{Fellow,~IEEE},
Abbas~Jamalipour,~\IEEEmembership{Fellow,~IEEE},
and Zhu~Han,~\IEEEmembership{Fellow,~IEEE}

\thanks{Manuscript received xxx. }

\thanks{Yaoqi~Yang is with the College of Communications Engineering, Army Engineering University of PLA, Nanjing 210000, China. (e-mail: yaoqi$\_$yang@yeah.net)}

\thanks{Hongyang Du and Dusit Niyato are with the College of Computing and Data Science, NTU, Singapore. (e-mail: hongyang001@e.ntu.edu.sg; dniyato@ntu.edu.sg)}

\thanks{Zehui Xiong is with the Information Systems Technology and Design Pillar, Singapore University of Technology and Design, Singapore. (e-mail: zehui$\_$xiong@sutd.edu.sg)}

\thanks{Abbas Jamalipour is with the School of Electrical and Computer Engineering, University of Sydney, Sydney, NSW 2006, Australia (e-mail:
a.jamalipour@ieee.org).}

\thanks{Zhu Han is with the Department of Electrical and Computer Engineering at the University of Houston, Houston, TX 77004, USA. (e-mail: hanzhu22@gmail.com)}
}

\maketitle

\begin{abstract}
The increasing demand for sensing, collecting, transmitting, and processing vast amounts of data poses significant challenges for resource-constrained mobile users, thereby impacting the performance of wireless networks. In this regard, from a case of mobile crowdsensing (MCS), we aim at leveraging attention mechanisms in machine learning approaches to provide solutions for building an effective, timely, and secure MCS. Specifically, we first evaluate potential combinations of attention mechanisms and MCS by introducing their preliminaries. Then, we present several emerging scenarios about how to integrate attention into MCS, including task allocation, incentive design, terminal recruitment, privacy preservation, data collection, and data transmission. Subsequently, we propose an attention-based framework to solve network optimization problems with multiple performance indicators in large-scale MCS. The designed case study have evaluated the effectiveness of the proposed framework. Finally, we outline important research directions for advancing attention-enabled MCS.
\end{abstract}


\IEEEpeerreviewmaketitle

\section{Introduction}

Attention mechanisms \footnote{In the paper we use ``attention" and ``attention mechanism" interchangeably.} can enable machine learning (ML) models to focus on specific parts of input data that are more relevant to the task at hand. Such an ability leads to advantages on model performance improvement, long sequence handling, and better interpretability. Such benefits have been evidently evaluated in various Artificial Intelligence (AI) applications, for example, 1) In the natural language processing (NLP) domain, attention has revolutionally enhanced performances in machine translation, text summarization, and sentiment analysis; 2) In the computer vision (CV) area, attention can enable image captioning and visual question answering, where the model needs to focus on specific parts to provide answers; 3) In the speech processing field, attention helps the ML model focus on relevant temporal parts of an audio signal, improving the accuracy of transcriptions. In summary, attention can offer significant benefits:
\begin{itemize}
  \item \emph{Handling long-distance dependencies of the input data}. Attention enables AI models to focus on different parts of the input data regardless of their position. This is particularly beneficial for complex structured inputs where relationships span across different segments of data, enhancing the retention of contextual information across longer sequences.
  \item \emph{Increasing the accuracy and fluency of generative outputs}. By focusing on important parts of the input when generating data, attention can improve the accuracy and quality of the output result. This is vital in tasks such as text generation, where maintaining grammatical correctness and context relevance is crucial for high-quality outputs.
  \item \emph{Enhancing the model flexibility and generalization performances}. Attention can help models to better generalize by learning which features to focus on, rather than just memorizing patterns specific to the training data. This ability enables AI models to perform better on varied and unseen data, thus enhancing their flexibility and utility across different scenarios and tasks.
  \item \emph{Improving the interpretability of the generative model}. By providing a mechanism to trace back to the parts of the input that were most influential in generating a specific output, attention can make AI model's decision-making process more transparent, aiding in understanding, debugging, and explaining the model's behavior, which is critical in deploying AI solutions responsibly.
\end{itemize}

Simultaneously, wireless networks serve as the foundation for connecting mobile users to a vast array of services, facilitating communication, data exchange, and information access. However, due to various service requirements, complex data interaction relationships, and open transmission environments, wireless networks need to efficiently and carefully deal with a large amount of data to make decisions on provided services, mobile users, and wireless data aspects. In this regard, it can be difficult to achieve high efficiency, satisfied quality, and strong security simultaneously within wireless networks. Specifically, wireless networks face challenges on fairness concerns of provided services, effectiveness management of mobile users, and timeliness guarantee of wireless data.


The widespread adoption of wireless networks in diverse mobile scenarios necessitates addressing above various challenges. While AI and machine learning have been widely adopted to address the challenges, they face certain limitations due to, for example, complex traffic data, complicated protocol optimizations, and limited generative network management solutions. Therefore, researchers have explored attention in wireless networks. To this end, we focus on mobile crowdsensing (MCS) as a compelling case to provide a comprehensive investigation and specific solutions, enhanced by attention.
MCS facilitates wireless data collection through crowdsourcing, where a service requester (SR) delegates sensing tasks to a service provider (SP). The SP recruits mobile workers (MWs), such as intelligent sensing terminals, to collect required data \cite{yang2024generative}. This approach effectively addresses concerns related to requirements for specialized sensing equipment, deployment of fixed sensing networks, and limitations of single sensing capability (i.e., data type or modality can be collected by terminals). However, MCS may face various challenges attributed to sensing tasks, MWs, and sensing data aspects. Inspired by aforementioned benefits, we investigate the use of attention to address concerns in MCS. Specifically, given the ability to focus on most important parts of input data, attention can potentially improve efficiency and accuracy of various tasks in MCS. Moreover, attention can offer insights to built fair, efficient, secure, and timely MCS in the following ways:
\begin{itemize}
\item \emph{Sensing task optimization}. Attention can dynamically allocate tasks by selectively processing historical and current environment information, ensuring better efficiency in task distribution. Additionally, by analyzing specific patterns in task completion and worker recruitment costs, attention can help design rational, fair, and cost-effective incentives, encouraging broader participation.
\item \emph{MWs management}. Attention can help to better recruit MWs based on the MWs' current  capabilities and historical performance, improving task completion rates and MWs' reward satisfaction. Furthermore, by monitoring sensing data traffic to detecting malicious attacks, attention can enhance privacy protections and create a more secure environment for MWs.
\item \emph{Sensing data enhancement}. Attention can effectively deal with complicated and dynamic sensing environments, facilitating better data collection policy learning and reducing interaction complexity among MWs and environments. Moreover, attention can help MCS to optimize wireless resource utilization for  data transmission, ensuring timeliness and usefulness of the sensing data.
\end{itemize}

In summary, attention excels at handling extensive input information by focusing on the most beneficial parts for further processing. Hence, integrating attention into MCS can improve task allocation, incentive design, MW recruitment, privacy protection, data collection and transmission within MCS. Specifically, the major contributions of this paper are summarized as follows:
\begin{itemize}
\item We first introduce preliminaries of MCS and attention. Next, we provide a comprehensive review to demonstrate necessities and benefits of integrating attention into MCS.
\item From the sensing task, MW, and sensing data perspectives, we investigate how to adopt attention in MCS, aiming at address fairness, effectiveness, and quality concerns.
\item We propose an attention-enabled network performance optimization framework for the large-scale MCS, and the case study has evaluated the effectiveness of the proposal.
\end{itemize}

\section{Preliminaries for Mobile Crowdsensing and attention mechanism}

\subsection{Preliminaries for Mobile Crowdsensing}

MCS represents a paradigm for leveraging the ubiquity of MWs (e.g., smart phones, watches, and vehicles) to gather and analyze sensing data. Generally, MCS is composed of an SR, an SP, and MWs. The SR first generates sensing data demand. Then, due to its limited sensing capacity, the SR publishes the demand to the SP. By recruiting MWs, the SP can obtain the required sensing data and send it to the SR. Following the above process, the data interactions among SR, SP, and MWs mainly include \cite{yang2024generative}:
\begin{itemize}
\item \emph{MW recruitment}. The SR recruits MWs capable of fulfilling SR's sensing demands.
\item \emph{Incentive design}. The SP designs incentive mechanisms to attract MWs to contribute sensing capabilities, in which the incentive can be monetary rewards, reputation credit, or other benefits.
\item \emph{Task allocation}. After the MWs are recruited, they are allocated to specific sensing tasks based on factors such as proximity, availability, and capability.
\item \emph{Data collection}. The MWs equipped with sensing modules collect required sensing data to finish the allocated tasks.
\item \emph{Data transmission}. After collecting the sensing data, the MWs send it to the SP over a wireless channel.
\item \emph{Privacy protection}. To safeguard personal information and sensitive data collected from MWs, privacy protection measures should be implemented.
\end{itemize}

By orchestrating these interactions effectively, MCS is able to utilize MWs to obtain required sensing data. However, due to the complex data inputs and interactions, heterogeneous sensing abilities and task demands, and open sensing environments and channel accesses, MCS faces challenges related to fairness, effectiveness, and data quality. In this context, it is essential to explore attention mechanisms to efficiently leverage extensive input information for constructing a sustainable, secure, and timely MCS.

\subsection{Preliminaries for Attention Mechanism}

\begin{table*}[!htb]
\centering
\caption{Summarization and comparison of different categories attention mechanisms.} \label{table}
\resizebox{0.9\width}{!}{
\begin{tabular}{c}
\includegraphics[width=15cm]{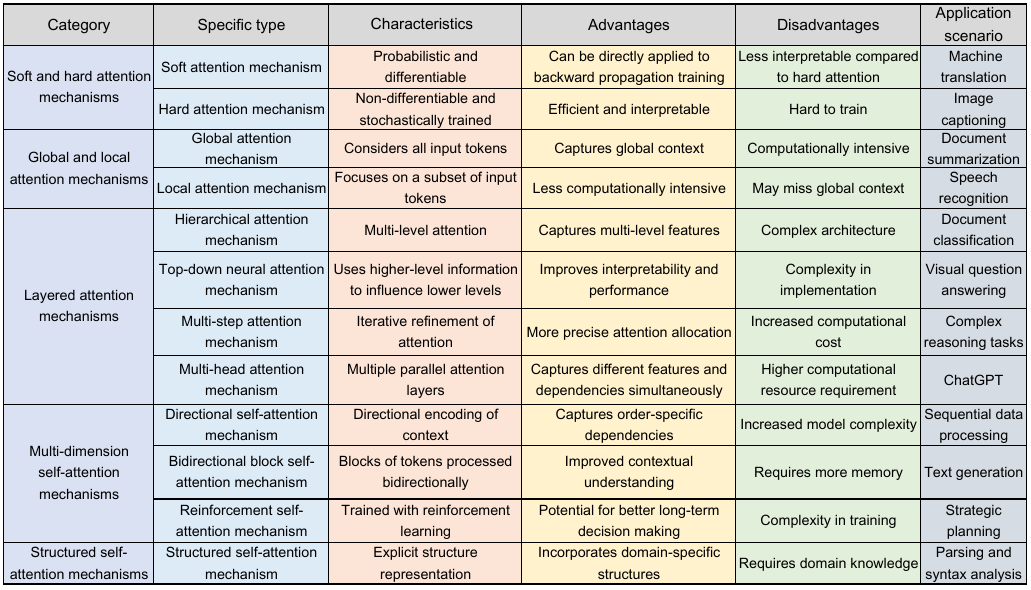}\\
\end{tabular}}
\end{table*}

An attention mechanism is typically adopted in neural networks, targeting at enhancing performance of machine learning models by allowing them to focus on relevant parts of the input data. As shown in Fig. 1(m), within the attention mechanism, context vectors indicate intermediate states or features extracted from the input sequence at each position (i.e., the index of each element in the input sequence data). Besides, a query refers to the element computing for the attention, key means the element compared against the query, and value indicates the element that are weighted and summed to produce the final output. Specifically, to implement attention mechanism, three steps should be executed \cite{hong2022balancing}:
\begin{enumerate}
\item \emph{Score calculation}. For a given query, the attention mechanism computes a score with each key. This score represents the relevance or importance of each key to the query. A common method for calculating the score is the dot product between the query and the key.
\item \emph{Softmax normalization}. The scores are then normalized using the softmax function to produce attention weights. These weights sum to one and represent the distribution of focus across all keys.
\item \emph{Weighted sum}. The final step involves multiplying attention weights by corresponding values and summing them together. This weighted sum is the output of the attention mechanism for the given query.
\end{enumerate}

Generally, the major categories of attention mechanisms include soft and hard attentions, local and global attentions, layered attentions, multi-dimension self-attentions, and structured self-attention (Table I and Fig. 1).
\begin{itemize}
\item \emph{Soft and hard attention mechanisms.} Soft attention assigns weights to all parts of the input, representing a weighted average of the input. These weights sum to one, indicating the relative importance of each part. In contrast, hard attention selects a subset of the input, focusing exclusively on these chosen parts. This involves making a discrete decision about the parts that are most relevant.
\item \emph{Global and local attention mechanisms.} Global attention considers all input elements when generating each output element. It recalculates context vectors for every output step, capturing global dependencies. Conversely, local Attention focuses on a smaller, dynamic subset of the input elements for each output time step, allowing for more efficient processing for long sequential data.
\item \emph{Layered attention mechanisms.} Layered attention is structured as layers, allowing the model to first attend to small local parts and progressively integrate information to focus on higher-level features. It mainly includes the hierarchical attention, top-down neural attention, multi-step attention, and multi-head attention.
\item \emph{Multi-dimension self-attention mechanisms.} Multi-dimension self-attention is an extension of self-attentions, which allows the model to attend not just to one dimension (such as time in sequences) but to multiple dimensions simultaneously. In this regard, rather than compute a scalar score for each component of the word embedding vector\footnote{A word embedding vector represents each word as a fixed-length vector, which captures the semantic information of the word. Each word embedding vector consists of multiple dimensions, each dimension representing a different feature or attribute of the word.}, multi-dimension self-attention computes a vector score for each component of the token embedding vector. Multi-dimension self-attention mainly includes the directional self-attention, bidirectional block self-attention, and reinforcement self-attention.
\item \emph{Structured self-attention mechanisms.} Structured self-attention considers structural dependencies of deep neural networks, transforming typical attention from a common soft choice, where each input is assigned a probability of being selected, to a new mechanism that models information about internal structures without disrupting end-to-end training. As a result, the structured self-attention models continuous, subsequence multiple-input selection instead of modelling single-input selection.
\end{itemize}

Recently, attention-enabled neural networks have been widely applied in the wireless networking systems mainly on three aspects:
1) Attention-based deep reinforcement learning (DRL) framework for strategies making, for example, \emph{Graph attention-based DRL for resource allocation \cite{ding2022resource}}. In vehicular communication networks, \cite{ding2022resource} aimed at maximizing the communication throughput by jointly optimizing spectrum and power resources. Subsequently, the study adopted a multi-agent DRL to solve above formulated problem. However, since the environments are complex and full of diverse dynamic information, it is difficult for vehicle agents to realize effective state information sensing to allocate resources. In this regard, by focusing on more relevant information (e.g., network status and channel conditions), attention can enhance efficiency of interaction and sensing of collaboration among vehicle agents. This can further significantly reduce signal overheads and improve communication performances. Experiment results showed that under the 1,060 Bytes payload, the proposed scheme can realize 54 Mb/s sum rate, while the meta-DRL scheme is only with 50 Mb/s sum rate.

2) Combing attention with discriminative AI for results prediction, classification and regression, for example, \emph{Sequence-to-sequence (seq2seq) attention-based convolutional neural networks (CNN)-long short-term memory (LSTM) for long time series forecasting \cite{wang2022long}}. To effectively predict long time series with rich information, CNN is first employed to extract multiscale features. This involves utilizing different convolution kernels to capture short-term features at various scales. Subsequently, the study integrated the seq2seq attention mechanism into the hidden state of the LSTM, enabling the generation of the output sequence by leveraging all previous input information. To this end, the proposed CNN-attention-LSTM scheme can accurately forecast long time series. Simulation results using real-world datasets demonstrate effectiveness of the proposed approach, achieving a Root Mean Square Error (RMSE) of 0.97, compared with the RMSE of 1achieved by the baseline LSTM without attention.

3) Integrating attention into generative AI for data generation and augmentation, for example, \emph{Slot attention-based Variational Autoencoder (VAE) for object-centric scene generation \cite{wang2023slot}}. When applying Variational Autoencoders (VAEs) to generate object-centric scene images, challenges arise in modeling complex object relationships within highly structured scenes and achieving high-quality sample generation. Slot attention offers a solution to these challenges by enabling object-centric representation inference. By integrating slot attention into a hierarchical VAE framework, the resulting model can simultaneously infer a global scene representation and object-centric slot representations. This allows for the capture of high-level scene structure and the embedding of individual object components, respectively. Numerical evaluations on the Arrow Room dataset demonstrate that the proposed approach achieves an S-Acc score of 0.94, significantly surpassing the baseline SIR scheme¡¯s score of 0.18. Such an attention application can provide image data to train specific wireless ML models, e.g., slot attention-based VAE can generate high-quality spectral maps of radiation sources for identification and localization\footnote{https://ietresearch.onlinelibrary.wiley.com/doi/10.1049/cmu2.12560?af=R}.


We summarize above attention mechanisms in Table I, where we present their categories, specific type, traits, advantages, disadvantages, and application scenarios. Moreover, we show their structures and typical networks in Fig. 1, where we highlight the implementation details of various attention mechanisms.

\begin{figure*}[!htb]
  \centering
  \includegraphics[width=15cm]{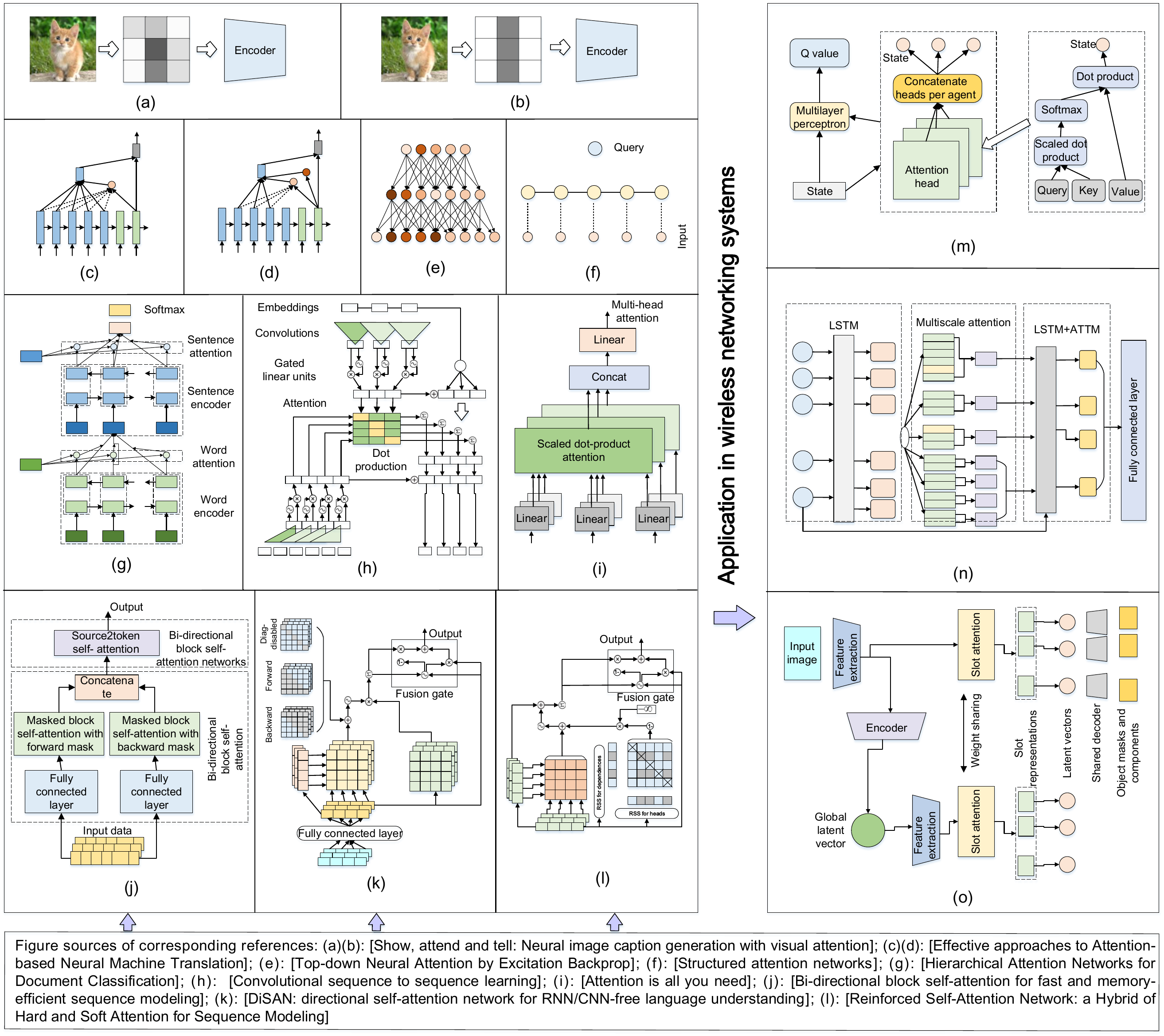}\\
  \caption{Structure of different attention mechanisms and applications in wireless networking systems.
  (a) Soft attention. 
  (b) Hard attention. 
  (c) Global attention. 
  (d) Local attention. 
  (e) Top-down neural attention. 
  (f) Structured self-attention. 
  (g) Hierarchical attention. 
  (h) Multi-step attention. 
  (i) Multi-head attention. 
  (j) Bidirectional block self-attention. 
  (k) Directional self-attention. 
  (l) Reinforcement self-attention. 
  (m) Graph attention-based DRL for resource allocation \cite{ding2022resource}. 
  (n) Seq2seq attention-based CNN-LSTM for long time series forecasting \cite{wang2022long}. 
  (o) Slot attention-based VAE for object-centric scene generation \cite{wang2023slot}. 
  }
  \label{structure}
\end{figure*}

\section{Integrating attention mechanisms into mobile crowdsensing}

\subsection{Research Focuses and Solutions}

\begin{figure*}[!htb]
  \centering
  \includegraphics[width=16cm]{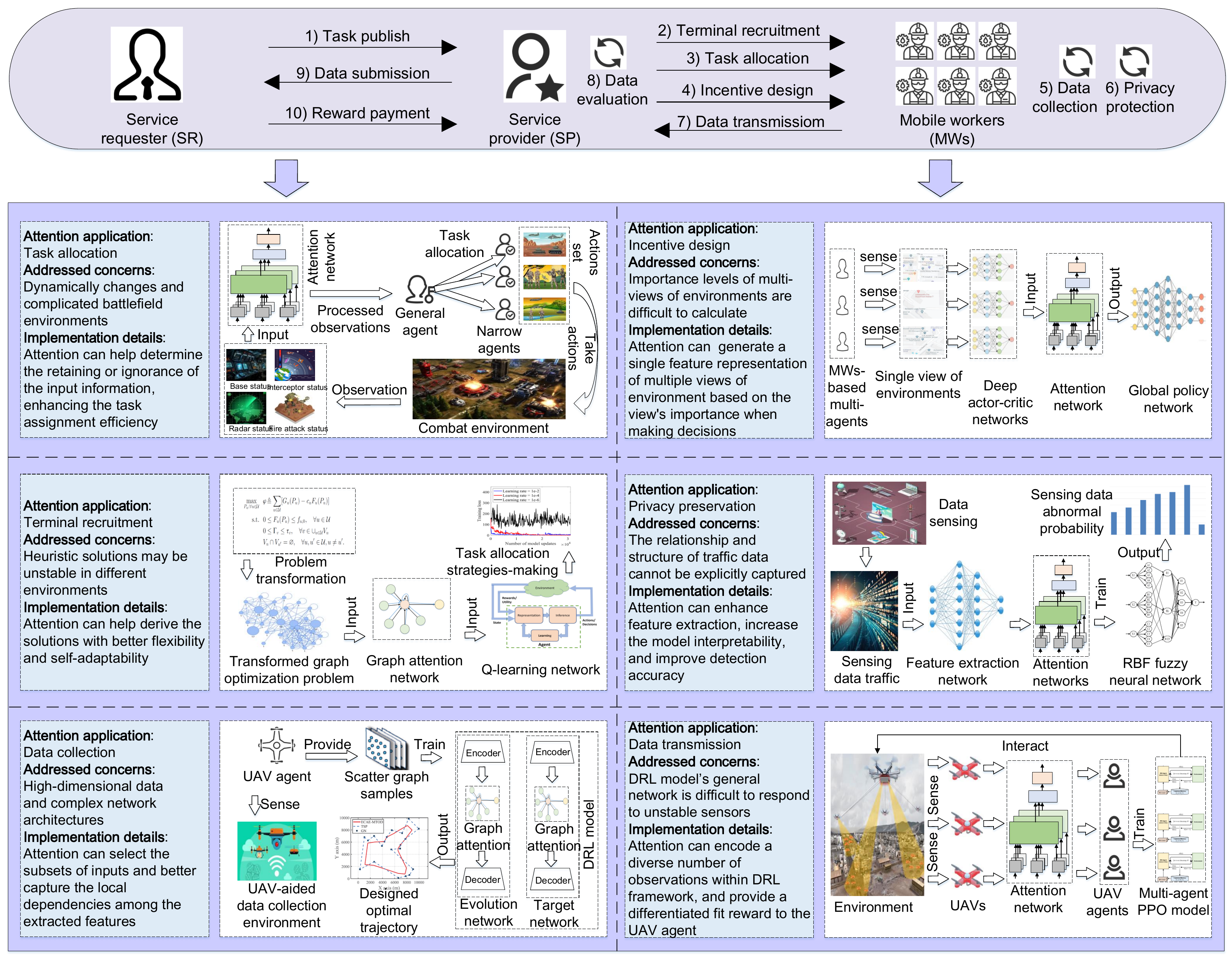}\\
  \caption{The role of attention mechanism from the sensing task, MWs, and sensing data aspects for MCS development and deployment, including task allocation, incentive design, MW recruitment, privacy preservation, data collection and transmission aspects.}
  \label{framework}
\end{figure*}

Fig. 2 presents a summary of how attention can enhance MCS. Attention can address fairness, effectiveness, security, and timeliness concerns from the sensing task, MWs, and sensing data aspects within MCS. Specifically, we summarize the primary research focuses and solutions as follows.

\subsubsection{Attention for MCS on sensing task aspect}\

\emph{Attention-enabled task allocation}. For the task assignment scenario in a ground-air confrontation system, the general agent is SR, and narrow agents are MWs \cite{liu2023task}. When allocating tasks to the narrow agents, the decision making can be based on a number of of global features supported by the general agent. However, dynamic and complex battlefields necessitate decision networks that consider both current and historical information, posing a challenge due to the volume of input data. Attention mechanisms can address this issue by selectively retaining or ignoring input information, thus improving task assignment efficiency. \cite{liu2023task} utilizes a Proximal Policy Optimization (PPO) framework for the task assignment, incorporating a multi-head attention mechanism (shown in Fig. 1(i)) within the general agent to process global situations. Simulation results show the proposed approach achieving a mean win ratio\footnote{The win ratio is proportion of wins achieved by a particular player or team out of the total number of games played. A high win ratio indicates efficient task allocation.} of 0.8, exceeding the 0.66 ratio of the baseline Alpha-C2 scheme.

\emph{Attention-enabled incentive design}. Typically, with a Stackelberg game framework, SP and MWs act as the leader and followers, respectively \cite{gu2020multiagent}. During competitive interactions, the SP first determines its own payoff. Then, the MWs employ a multi-agent DRL to adjust their strategies. However, the agents for MWs may suffer from partial observability of the environment, only exploring a single view. Creating multiple agents allows for observing multiple views of the environment.
While multiple views can benefit learning performance, accurately calculating their importance levels is challenging. Attention can precisely determine the importance of each view. For example, \cite{barati2019actor} uses attention (shown in Fig. 1(i)) to generate a single feature representation from multiple views based on their importance during decision-making. Numerical results demonstrate that the proposed approach achieves the data quality-oriented reward of 4,100 compared with 2,000 and 3,000 for the baseline Asynchronous Advantage Actor-Critic (A3C) and Deep Deterministic Policy Gradient (DDPG) schemes, respectively.

\subsubsection{Attention for MCS on MWs aspect}\

\emph{Attention-enabled MW recruitment}.
Within a terminal recruitment scheme, both SP and MWs can make an agreement to successfully finish sensing tasks.
Generally, the MW recruitment problem is NP-hard, and heuristic solutions could suffer from unstable performance during dynamics in different environments. In this regard, attention can help the SP and MWs to derive the solutions with better flexibility and self-adaptability. For example, \cite{xu2023intelligent} first models the
NP-hard terminal recruitment problem as a graph-based optimization problem. Then, through integrating a graph attention network into DRL, the MWs-oriented agents can more efficiently explore the solution space of the targeted problem, realizing better solution performance. Numerical results have demonstrated that when the number of tasks is 60, the proposed attention-based approach can achieve the profit of 8,000 units, while the baselines Greedy and deep Q-learning (DQN) schemes are only with around 2,500 and 4,300 units.

\emph{Attention-enabled privacy preservation}.
MCS raises privacy concerns as collected data may contain sensitive information about MWs \cite{yang2024generative}. Protecting MW privacy is crucial, but existing traffic detection approaches for privacy preservation are computationally expensive and struggle to capture the structure of large amounts of sensing data. Attention can enhance the input traffic data processing efficiency and improve model detection performance. For example, \cite{liu2023abnormal} introduces an attention to the fuzzy neural networks-based abnormal traffic detection scheme. Specifically, fuzzy rules can simulate factors relationships in the input data. By combining incremental and batch fuzzy neural network training methods, the detection network is trained comprehensively. Moreover, attention (shown in Fig. 1(i)) can boost feature extraction, increase the model interpretability, and improve detection accuracy. Numerical results have shown that the proposed scheme can realize the detection accuracy of 0.9995, while  decision tree and support vector machine (SVM) approaches achieve only 0.7884 and 0.8575, respectively.

\subsubsection{Attention for MCS on sensing data aspect}\

\emph{Attention-enabled data collection}. Unmanned Aerial Vehicles (UAVs) can serve as MWs to collect sensing data within MCS. However, due to limited energy supply of UAVs, it is important to minimize UAV's and sensor's energy consumption as well as maintaining data acquisition efficiency by optimizing the UAV trajectory. However, when optimizing the trajectory with DRL approach in the solution space, the computational cost and solution performance can be excessive owing to the high-dimensional data and complex network architectures. In this regard, attention can address such concerns by selecting subsets of inputs and better capture the local dependencies among extracted features. For example, \cite{zhu2023uav} transforms the input data as the graph structural data, and then integrated the graph attention into the evaluation and target networks of DRL model. After training the attention-enabled DRL model, an optimal UAV trajectory can be derived from the outputs. Simulation results demonstrate that, under identical packet reception rate settings, the proposed approach achieves 27.1\% reduction in total energy consumption compared with the Successive Convex Approximation (SCA)-based scheme for both UAVs and sensors.

\emph{Attention-enabled data transmission}. In UAV-enabled MCS, optimizing resource allocation for data transmission from ground sensors is crucial. Traditional approaches using PPO-based DRL may struggle with dynamic changes in the number of active sensors.
Since the input size of DRL model's general network is fixed, it is difficult to effectively respond to the changing sensor information. In this regard, attention can alleviate the issue of varying number of sensors in the input of the PPO-based DRL \cite{park2023joint}. Specifically, by adding multi-head attention in front of the input layer of the DRL model, a learning network is proposed to handle a varying number of active sensors \cite{park2023joint}. On this basis, attention (shown in Fig. 1(i)) can encode a diverse number of observations within the DRL framework and provide a well-designed reward to the UAV agent. Simulation experiments have evaluated that the proposed scheme can achieve 3.53\% and 19.74\% achievable rate improvement compared with fair and random resource allocation baseline schemes.

\subsection{Lessons Learned }

In summary, attention can significantly enhance input data processing efficiency, improve feature extraction within neural network models, and improve model interpretability. These advantages contribute significantly to MCS by improving fairness, effectiveness, sustainability, security, and timeliness. However, despite its benefits, attention-enabled MCS faces certain challenges due to limitations inherent to attention mechanisms.
\begin{enumerate}
\item \emph{Overfitting risk}. Attention may focus too heavily on specific data features that can lead to overfitting, where the model learns noise or irrelevant patterns, hindering the ability to generalize to new data.
\item \emph{Hyperparameters tuning}. Attention introduces additional hyperparameters that require careful tuning for optimal performance, often requiring significant experience and experimentation.
\item \emph{Limited sample efficiency}. Attention may not be efficient with limited data, struggling to extract meaningful patterns and relationships when data is scarce or expensive to obtain.
\end{enumerate}

\section{Attention-enabled multiple performance indicators optimization for large-scale mobile crowdsensing}

\subsection{Research Motivations}

To meet the requirements of MCS, multiple performance indicators need to be jointly optimized. While multi-objective evolutionary algorithms can effectively solve such problems \cite{hong2022balancing}, complex sensing tasks requiring large datasets, and numerous MWs introduce scalability and complexity challenges. This leads to exponential growth in decision variables and search spaces, resulting in the curse of dimensionality for traditional multi-objective evolutionary algorithms.
To address this, balancing exploration and exploitation of decision variables is crucial, as they contribute differently to the optimization process \cite{hong2022balancing}. Attention mechanisms, having the ability to focus on specific areas, can enhance processing efficiency and improve performance. Therefore, we integrate attention into multi-objective evolutionary algorithms for optimizing multiple performance indicators in large-scale MCS. Attention can assign a unique weight to each decision variable, enabling a balanced exploration and exploitation at the variable level. This effectively addresses the curse of dimensionality and improves optimization results \cite{hong2022balancing}.

\subsection{Proposed Framework}

\begin{figure*}[!htb]
  \centering
  \includegraphics[width=14cm]{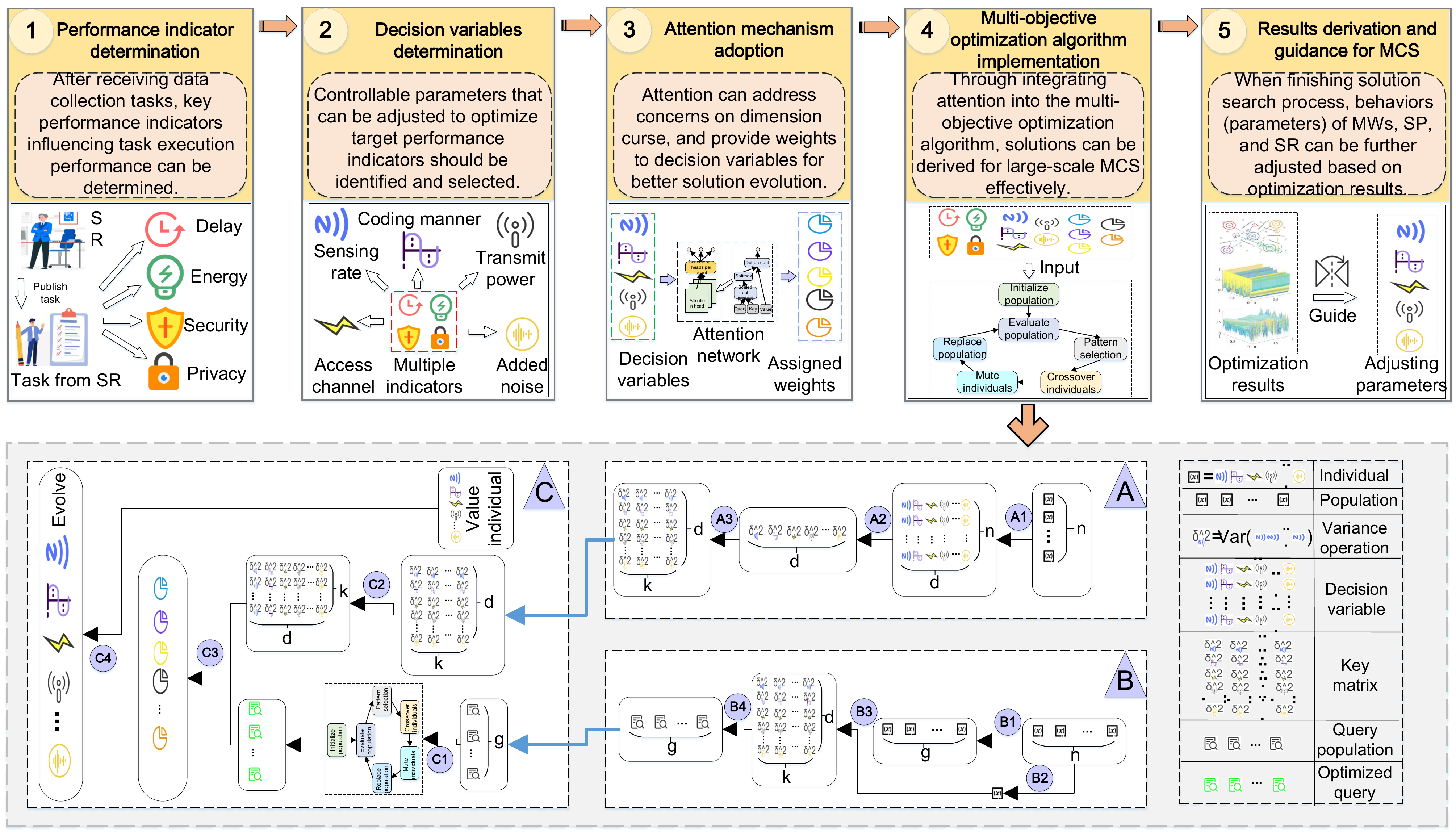}\\
  \caption{The framework of attention mechanism-based optimization with multiple performance indicators in large-scale MCS. In the multi-object optimization algorithm implementation step, three stages are included. Specifically, stage A determines the key matrix. Stage B calculates the query population, and Stage C implements the large-scale multi-objective optimization algorithm with attention mechanism. $n$, $d$, $k$, and $g$ represent the population's dimension, the individuals' number, the query's dimension, and random selected individuals' number, respectively.
  }
  \label{framework12}
\end{figure*}

With the help of an attention mechanism, multiple performances can be jointly optimized in large-scale MCS. Accordingly, as shown in Fig. 3, following a similar analysis in \cite{hong2022balancing}, we propose a framework for realizing attention-based multi-performance-indicator optimization in large-scale MCS, which is implemented with following five steps.

\emph{Step 1: Performance indicator determination}. After receiving a specific sensing task, a comprehensive analysis should be performed to identify key performance indicators (KPIs) for task-oriented, large-scale MCS. For example, in a real-time monitoring-oriented MCS scenario involving ground sensors, KPIs such as Age of Information (AoI), sensor sensing energy, and transmission energy consumption should be considered for joint optimization to ensure sustainability and timeliness.

\emph{Step 2: Decision variables determination}. Based on the identified performance indicators, the decision variables influencing those indicators should be determined. In essence, controllable parameters within the large-scale MCS that can be adjusted to optimize target KPIs should be identified and selected. Then, decision variables influencing the KIPs of energy consumption and transmission delay could include sensing rate, transmission power, and allocated bandwidth.

\emph{Step 3: Attention mechanism adoption}. A large scale of MCS can lead to an exponential increase in the dimension of decision variables and solution spaces. To address this dimensionality challenge, attention mechanisms are employed. Specifically, the key matrix within the attention mechanism is derived from the variance of the current solution space across different decision variables, enabling decision variable classification and extraction. A query is then calculated by multiplying the decision vector with the key matrix. The resulting attention vector is obtained by multiplying the query with the transpose of the key. This attention vector guides the evolution of the solution, effectively addressing the dimensionality problem.

\emph{Step 4: Multi-objective evolutionary algorithm implementation}. As explained in Step 4 of Fig. 3, when implementing a multi-objective optimization algorithm, an individual represents a candidate solution to the problem, consisting of a set of decision variables. A population is a collection of individuals representing the current search space in each generation. Fitness measures an individual¡¯s performance based on the objectives, and selection involves choosing better-performing individuals to create the next generation. Crossover combines characteristics from multiple individuals to generate new solutions, while mutation introduces random changes. The new individuals are integrated into the current population through an update process, forming a new population. Diversity maintenance aims to prevent premature convergence and ensure adequate exploration of the search space by preserving diversity in the population. Within the considered framework, the optimization performance indicators can be regarded as fitness, decision variables constitute the individuals and population, and attention is utilized to assign weights to decision variables for enhanced solution derivation. Moreover, crossover and mutation are the evolutionary operations, while diversity maintenance guides the derivation of optimal solutions.
To be detailed, stage A determines the key matrix. A1 can determine all the decision variables from population. A2 can calculate column variance from all decision variables to obtain the variance vector. A3 can construct the key matrix based on variance vector, where $k$ is a hyperparameter indicating dimension of query.
Besides, stage B calculates the query population.
B1 aims at randomly selecting $g$ solutions from population. B2 can determine the value individual based on largest crowding distance on the first front. B3 refers to multiplying each decision vector and value individual by key matrix. Then, by calculating the ratio of each dimension of above two obtained matrices, B4 can determine the query's elements.
Moreover, stage C implements the proposed attention-based algorithm. C1 optimizes the query results through the evolution algorithm. C2 conducts the transpose operation for the key matrix. C3 can obtain the attention vector by multiplying optimized query and key's transpose matrix. By multiplying attention vector and value individual, then adding the results to population, C4 can realize the individual evolution.

\emph{Step 5: Results derivation and guidance for MCS}. After solving the multi-objective optimization problems for large-scale MCS, the derived solutions can guide the behavior of MWs, SP, and SR. For instance, based on the optimization results, the sensing rate and transmission power strategies of MWs can be adjusted to achieve better data freshness and energy efficiency.

\subsection{Case Study and Results Analysis}

\begin{figure*}[!htb]
  \centering
  \includegraphics[width=11cm]{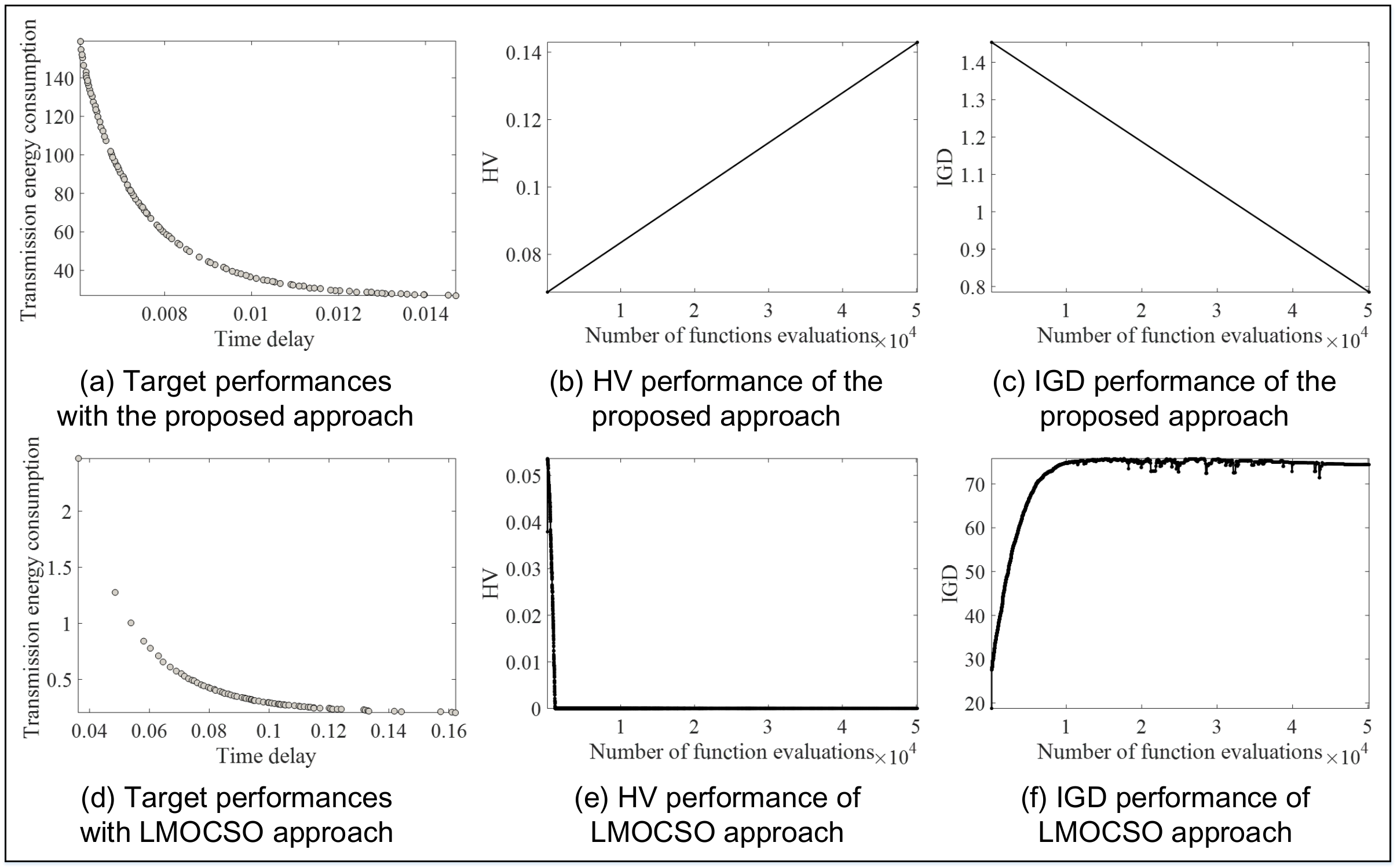}\\
  \caption{Performance evaluation for the proposed framework. (a) Target performances with the proposed approach. (b) HV performance of the proposed approach. (c) IGD performance of the proposed approach. (d) Target performances with the LMOCSO approach. (e) HV performance of the LMOCSO approach. (f) IGD performance of the LMOCSO approach. }
  \label{structure666}
\end{figure*}

We consider a large-scale MCS scenario involving UAV-aided sensing data collection, which can be applied in environmental monitoring, target reconnaissance, and low-altitude warning tasks \cite{yang2023mean}. In this scenario, the SR publishes sensing tasks to the SP, which is a UAV (tethered balloon in this context). The UAV recruits ground sensors to collect the required data. After the data collection, the sensors transmit the data to the UAV for further processing. To ensure efficient and timely data transmission, the target performance indicators to be optimized are the transmission energy consumption of ground sensors and transmission delay \cite{luan2020energy}. The decision variables are the individual transmission powers of different ground sensors.

To evaluate the effectiveness of the proposed framework, we conduct simulation experiments for the above considered scenario with parameter settings in \cite{hong2022balancing} and \cite{luan2020energy}. All the experiments are conducted on PlatEMO platform\footnote{https://github.com/BIMK/PlatEMO}, running with Matlab R2023b on a laptop with Intel(R) Core(TM) i7-11700 CPU @ 1.60GHz, NVIDIA GeForce RTX 3060 GPU, 32.0 GB memory and Windows 11. The code of this case study is available on github\footnote{https://github.com/Yaoqi-Yang97/Attention4MCS}. When the number of MWs is set as 300, and the maximum number of functions evaluation is 50,000, Figs. 4(a) and (d) show the evolution results of the proposed approach and baseline, large-scale multi-objective optimization based on a competitive swarm optimizer (LMOCSO) scheme \cite{tian2019efficient}.
To demonstrate the effectiveness of the proposal, we adopt two indicators, Hypervolume (HV) and Inverted Generational Distance (IGD), to evaluate the performance of optimized multiple indicators. On the one hand, since HV refers to the volume of the multi-dimensional space from the reference point to all solutions, the larger the HV, the closer the solution set is to the Pareto front, and the better the diversity of the solutions. In this regard, from Figs. 4 (b) and (e), the proposed approach can achieve a larger HV value, evaluating its superiority. On the other hand, IGD is the average minimum distance from each point in the reference Pareto front to the solution set. A smaller IGD value indicates that the solution set is closer to the true Pareto front and has better coverage. On this basis, as shown in Figs. 4 (c) and (f), the proposed approach can achieve a smaller IGD value, demonstrating its advantages.

\section{Future research directions}

\subsection{Designing Transfer Learning-oriented Attention}

Designing transfer learning-oriented attention mechanisms can enable models to leverage pre-trained knowledge more effectively across different MCS tasks. For example, few-shot learning can be adopted to enhance attention mechanisms to work effectively with limited data, facilitating rapid adaptation to new tasks with minimal training.

\subsection{Lightweighting the Attention Model to Extend Applications}

In MCS, the implementation environment of attention is resource-constrained due to limited resource supply of mobile devices. Therefore, lightweighting attention models is crucial.  Advanced compression techniques, such as pruning, quantization, and knowledge distillation, can be used to lightweight the attention model without compromising performances.

\subsection{Investigating Context-aware Attention Mechanisms}

Context-aware attention mechanisms can significantly enhance the relevance and accuracy of data processing in MCS. Such a research can develop attention mechanisms that dynamically adjust focus based on contextual information, e.g., location, time, and user activity, leading to more precise and timely data analysis and improving MCS overall performances.

\section{Conclusion}

In this paper, we investigated the combination of attention with wireless networks, where MCS was set as a specific case. To be detailed, we first comprehensively introduced the preliminaries of MCS and attention, disclosing the motivations and benefits of integrating attention into MCS. Then, we presented several promising attention applications to MCS, mainly including task allocation, incentive design, terminal recruitment, privacy preservation, data collection, and data transmission aspects. In addition, their possible challenges of such integrations were also discussed. Moreover, we proposed a framework of attention mechanism-based multiple performance indicators optimization in large-scale MCS. The case study and simulation results have fully evaluated the effectiveness of the proposal framework. Finally, several potential research directions were highlighted for future work.

\bibliography{ref}{}
\bibliographystyle{IEEEtran}

\end{document}